# Diabatic Hamiltonian Construction in van der Waals heterostructure complexes

*Yu Xie[a,b], Huijuan Sun[c], Qijing Zheng[d,e], Jin Zhao[d,e], Hao Ren*[*f] and Zhenggang Lan*[*a,b]*

[a] SCNU Environmental Research Institute, Guangdong Provincial Key Laboratory of Chemical Pollution and Environmental Safety & MOE Key Laboratory of Theoretical Chemistry of Environment, South China Normal University, Guangzhou 510006, China

[b] School of Environment, South China Normal University, University Town, Guangzhou 510006, China

[c] College of Physics, Qingdao University, Qingdao, Shandong 266071, China

[d] Department of Physics, and Hefei National Laboratory for Physical Sciences at Microscale, University of Science and Technology of China, Hefei, Anhui 230026, China

[e] International Center for Quantum Design of Functional Materials (ICQD), CAS Key Laboratory of Strongly-Coupled Quantum Matter Physics, and Synergetic Innovation Center of Quantum Information & Quantum Physics, University of Science and Technology of China, Hefei, Anhui 230026, China

[f] School of Materials Science and Engineering, China University of Petroleum (East China), Qingdao, Shandong, 266580, China

# ABSTRACT

A diabatization method is developed for the approximated description of the photoinduced charge separation/transfer processes in the van der Waals (vdW) heterostructure complex, which is based on the wavefunction projection approach using a plane wave basis set in the framework of the single-particle picture. We build the diabatic Hamiltonian for the description of the interlayer photoinduced hole-transfer process of the two-dimensional vdW $MoS_2/WS_2$ heterostructure complexes. The diabatic Hamiltonian gives the energies of the localized valence band states (located at $MoS_2$ and valence band states (located at $WS_2$), as well as the couplings between them. The wavefunction projection method provides a practical and reasonable approach to construct the diabatic model in the description of photoinduced charge transfer processes in the vdW heterostructure complexes.

**TOC**

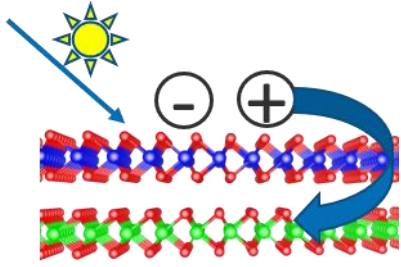

$$\begin{bmatrix} H_1^a & 0 \\ 0 & H_2^a \end{bmatrix} \Rightarrow \begin{bmatrix} H_{11}^d & H_{12}^d \\ H_{21}^d & H_{22}^d \end{bmatrix}$$

Photo-induced charge separation/transfer processes widely exist in van der Waals (vdW) heterostructure complexes. [1-4] Due to their importance, considerable efforts were made to investigate the photoinduced charge transfer dynamics of these materials from the theoretical point of view.[3, 5-11] For example, transition-metal dichalcogenide (TMD) van der Waals heterostructures $MoS_2/WS_2$ attracted tremendous interests due to their potential applicability as optoelectronic and solar energy conversion devices. [1-4, 12-15] By using photoluminescence and femtosecond pump-probe measurements, Hong et al. [12] found that the hole transfer in $MoS_2/WS_2$ was at the femtosecond scale (50 fs), indicating it is a promising candidate for optoelectronic and photovoltaic applications. The ultrafast charge transfer at TMD heterostructure interface has been further convinced by other experiments. [13-15] This process has also been investigated by different theoretical simulations [3, 5-11]. For example, Zheng et al.[6] used ab initio nonadiabatic molecular dynamics based on time-dependent Kohn-Sham equation and surface hopping method to show that the ultrafast interlayer hole transfer is strongly promoted by an adiabatic mechanism through phonon excitation occurring on 20 fs, which is consistent with the experimental results. Wang et al. [5] used Ehrenfest dynamics based on real-time time-dependent density functional theory (TDDFT) to investigate this process and found that the charge transfer occurs at 100 fs, accompanying with charge oscillation between the two layers. [5] All of these simulations provide the deep physical insight of the photoinduced hole transfer dynamics, while it is still important to develop other theoretical framework due to the deficiencies of the surface hopping and Ehrenfest dynamics.

In the analysis of charge transfer processes, the diabatic representation is a very attractive view that provides an intuitive physical picture with the diabatic states with localized charge distribution. [16] However, most electronic-structure calculations are performed in the adiabatic representation. Thus, it would be helpful to construct a diabatic model from ab initio electronic structures to describe the of photoinduced charge transfer processes in vdW heterostructure complexes.

In the past several years, several diabatization methods have been developed in the quantum chemistry community.[17-38] Many of them were mainly designed to build the diabatic wavefunctions for molecular systems when the adiabatic wavefunctions are available.[17, 19-25, 39-41] Alternatively, it is also possible to build the diabatic states directly by assuming the smooth property constrain of the diabatic electronic wavefunction.[26, 28-32, 42-44]

Although many diabatic methods have been well established, it is still not easy to construct the diabatic Hamiltonian for the large or extended systems, if many-electron wavefunctions that are the linear combination of the slater determinants are considered. For the treatment of the vdW heterostructure complexes, it is quite normal to employ approximated wavefunctions generated by electronic structure calculations. For instance, the single-particle picture was often adapted in the widely-used density functional theory calculations. In the simplified physical picture,[33, 45] the single-particle picture was adopted to describe the charge dynamics in the extended systems, where the many-body dynamics is simplified by the hole/electron motion from the donor states (orbitals) to the acceptor states (orbitals), with the electron-hole interaction neglected. These approximated and practical approaches provided useful understanding of the charge dynamics of extended systems. In this sense, it is also interesting to establish the diabatization procedure for such large systems. Some efforts were also made to build the diabatic models for large or extended systems, including the extended-Hückel approach [45], energy-broadening analysis [34, 38], projection-operator approach and block diagonalization of Fock matrix [33, 46], constrained density functional theory (CDFT) [32, 47-48], molecular orbital based fragmentation approaches[49-50] and so on. Particularly, these approaches were implemented to deal with the extended systems. For instance, CDFT was implemented in the CP2K,[51-52] SIESTA[53] and Quantum Espresso[54] packages. Some efforts were made to combine CDFT with the electronic structure calculations with the plane-wave basis set within VASP [55] and GPAW [56] packages.

Alternatively, it is also possible to construct the diabatization approaches by performing the matrix/vector transformation based on the output information in the electronic structure calculation performed with electronic structure packages. In this way, we do not need to modify the code of electronic structure packages. Instead, the whole diabatization procedure can be performed after the electronic structure calculations. When the diabatization method is constructed, it is also possible to develop the useful interface to different packages. This allows more flexible usage of the diabatization no matter what electronic structure packages are used. One of such diabatization methods is the projection-operator approach with block diagonalization of Fock matrix, which attracted considerable attention and has been employed to charge transfer across molecule/Metal ( or molecule/semiconductor) interfaces.[46, 57-58] Many implementations the projection approach with the block diabatization of the Fock matrix[46] reply on the localized atom-centered basis sets, because it is very clear that these method should be suitable when the basis functions are localized ones. However, a plane-wave basis set is usually used to expand the electronic structures of extended systems. In this case, the implementation of the previously mentioned diabatization schemes would be challenging, because all basis functions are completely delocalized. In this sense, it is very important to presume a suitable theoretical method to construct the diabatic Hamiltonian based on the electronic structure with the plane wave basis set.

The wavefunction projection method[19-20, 23, 59-64] is a practical and effective approach to construct the diabatic states. In this framework, the reference diabatic states are pre-defined. Then the projection of the adiabatic states at any calculated structure to the diabatic reference states defines the adiabatic-to-diabatic transformation, which is required to construct the diabatic Hamiltonian. In the context of photoinduced energy or electron transfer, the wavefunction of a specified electronic excited state is generally written as the linear combination of Slater determinants. This diabatization approach is widely employed to study the photoinduced exciton dynamics of organic photovoltaic systems. [65-66] Recently, Subotnik and coworkers proposed several protocols (including the projection approach) to construct the localized diabatic excitonic states in molecular crystal systems.[20]

In this work, we extended the wavefunction projection method to construct the diabatic Hamiltonian for vdW heterostructure complexes within the single-particle picture. For broad applications, the projection method is combined with the widely-used electronic structure calculations with plane wave basis sets. We employed such diabatic Hamiltonian to discuss the hole transfer in the $MoS_2/WS_2$ van der Waals heterostructure. After the construction of the diabatic Hamiltonian, it is possible to obtain the hole-transfer relevant states localized on each layer and the couplings between these states. This work provides a practical approach to construct the diabatic Hamiltonian of the photoinduced hole transfer in the $MoS_2/WS_2$ complex, thus, this may serve as a starting point for the future nonadiabatic dynamics of extended systems by using more rigorous theoretical approaches, such as multilayer multiconfiguration time-dependent Hatree (ML-MCTDH) [67], the hierarchy equation of motion (HEOM)[68-70] and so on.

When the diabatic picture is used to describe the electron transfer (ET) process, the photo-generated electron-donor states (denoted as $|\psi_d\rangle$) are coupled to electron-acceptor states (denoted as $|\psi_a\rangle$). The electronic Hamiltonian in the diabatic representation reads as follows:

$$H = \sum_d |\psi_d\rangle V_{dd} \langle\psi_d| + \sum_a |\psi_a\rangle V_{aa} \langle\psi_a| + \sum_{da} \left(|\psi_d\rangle V_{da} \langle\psi_a| + c.c\right). \quad (1)$$

where $V_{dd}$ and $V_{aa}$ denote the energies of the electronic donor and acceptor states, respectively. The off-diagonal matrix elements $V_{da}$ ($V_{ad}$) characterize the donor-acceptor electronic coupling.

Let us assume that we only consider the arbitrary vdW complex with two layers labelled as **Y** and **Z**, respectively, where Y is responsible for the photoabsorption. In such complicated systems, the one-particle electronic wave functions, in particular some frontier states, were often employed for the qualitative understanding of electron/hole processes. When the photoexcitation of the **Y** layer creates a hole in the valence band (VB) and an electron in the conduction band (CB) in the **Y** layer, the charge transfer may happen between two stacked layers of the whole complex, see Scheme 1. For both of them, we can understand them in the same theoretical framework in the diabatic representation within the current single particle picture. For the electron transfer [see Scheme 1(a)] in the vdW complex, it corresponds to the motion of an electron from the CB of the **Y** layer to the CB band of the **Z** layer. In this sense, the electron donor (D) and the electron acceptor (A) are defined as **Y_CB** and **Z_CB**, respectively. We can also understand the hole transfer [see Scheme 1(b)] by considering the electron motion. The hole transfer means that the hole moves from **Y_VB** to **Z_VB**. This corresponds to the electron transfer from **Z_VB** to **Y_VB**. Then the electron donor and acceptor become **Z_VB** and **Y_VB**, respectively. Thus, both of the electron and hole transfer can be described by the above electron transfer Hamiltonian if we define the suitable electron donor and acceptor. According to this idea, the diabatic Hamiltonian can be constructed by the projection of the total electronic wavefunctions of the complexes to the electronic wavefunctions of each component. Next, we try to outline the essential ideas of the diabatization approach.

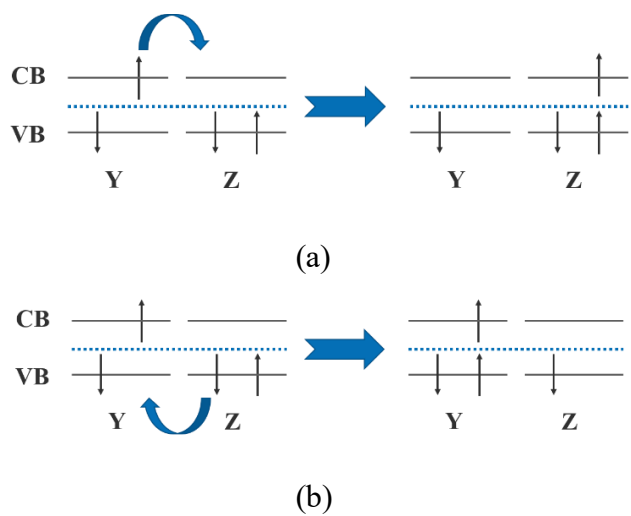

(a)

(b)

Scheme 1. Schematic diagram for electron transfer (a) and hole transfer (b) between **Y**

and **Z** layer in **Y/Z** heterostructure complex.

For the two-dimensional van der Waals heterostructures (such as $MoS_2/WS_2$), we separate the whole complex into two parts: the $MoS_2$ part and the $WS_2$ part. Please notice that we wish to study the $MoS_2 \rightarrow WS_2$ hole transfer in this system. In the electron motion picture, $WS_2$ is the electron donor and $MoS_2$ is the electron acceptor.

First the electronic wavefunction of two components were computed individually, and then we combined two wavefunctions to construct the reference states $\left|\Psi_i^{(ref)}\right\rangle$ of the whole complex. These states are localized on the two isolated parts, which can be taken as reference diabatic states.

Second, we compute the electronic wavefunction of the whole van der Waals heterostructure complex to obtain the adiabatic states $\left|\Psi_i^{(a)}\right\rangle$. The adiabatic-to-diabatic transformation matrix **T** is defined as the matrix can transform the adiabatic states $\left|\Psi_i^{(a)}\right\rangle$ to the diabatic states $\left|\Psi_i^{(d)}\right\rangle$,[19, 23, 64] i.e.,

$$\left|\Psi_j^{(a)}\right\rangle = \sum_i t_{ij} \left|\Psi_i^{(d)}\right\rangle. \qquad (2)$$

where $t_{ij}$ is the element of the transformation matrix **T**. By projecting the electronic states of the whole complex to the reference states $\left|\Psi_i^{(ref)}\right\rangle$,

$$\tilde{t}_{ij} = \left\langle \Psi_i^{(ref)} \middle| \Psi_j^{(a)} \right\rangle, \qquad (3)$$

we can obtain a matrix $\tilde{\mathbf{T}}$ with elements $\tilde{t}_{ij}$, which is defined as the projection matrix. In order to avoid the influence of the states outside the projection basis space, the orthogonalization is necessary. Thus, we finally build the adiabatic-to-diabatic transformation matrix **T** by applying the orthogonalization using

$$\mathbf{T} = \tilde{\mathbf{T}}\left(\tilde{\mathbf{T}}^\dagger \tilde{\mathbf{T}}\right)^{-1/2}. \qquad (4)$$

The diabatic Hamiltonian matrix is easily obtained as

$$\mathbf{V}_{el}^{(d)} = \mathbf{T}\mathbf{V}_{el}^{(a)}\mathbf{T}^\dagger, \qquad (5)$$

where $\mathbf{V}_{el}^{(d)}$ is the diabatic electronic Hamiltonian matrix and $\mathbf{V}_{el}^{(a)}$ is the adiabatic electronic Hamiltonian matrix.

Next the problem is how to calculate the wavefunction overlap in Eq. (3) when the plane-wave basis is employed to treat the periodic systems. The projector augmented-wave method (PAW) is used in calculations. The details on PAW method can be found in previous references. [71-72] We only provide a brief description here. In the PAW approach, the all-electron wave function $\left|\Psi_i^{(a)}\right\rangle$ can be written as the linear transform of the pseudo wave function $\left|\Psi_i^{(ps)}\right\rangle$,

$$\left|\Psi_i^{(a)}\right\rangle = \left|\Psi_i^{(ps)}\right\rangle + \sum_m \left(\left|\phi_m\right\rangle - \left|\phi_m^{(ps)}\right\rangle\right)\left\langle p_m^{(ps)}\middle|\Psi_i^{(ps)}\right\rangle, \qquad (6)$$

where $\left|\phi_m\right\rangle$ are the all-electron partial waves that are obtained from the reference atom. Please notice that here "all-electron wave function" is a widely used name that refers to the real electronic wavefunction instead of the pseudo wave function. This all-electron wavefunction is in the single-particle picture (orbital), instead of the many-electron picture (linear combination of slater determinants). $\left|\phi_m^{(ps)}\right\rangle$ are pseudo partial waves that are the same as all-electron partial waves outside a core radius and match continuously onto $\left|\phi_m\right\rangle$ inside the core radius. $\left|p_n^{(ps)}\right\rangle$ are the projector functions. The pseudo wave functions must fulfill the following orthogonality condition:

$$\left\langle \Psi_i^{(ps)}\middle|1 + \sum_{m,n}\left|p_m^{(ps)}\right\rangle q_{mn}\left\langle p_n^{(ps)}\right|\middle|\Psi_j^{(ps)}\right\rangle = \left\langle \Psi_i^{(a)}\middle|\Psi_j^{(a)}\right\rangle = \delta_{ij}, \qquad (7)$$

where

$$q_{mn} = \left\langle \phi_m\middle|\phi_n\right\rangle - \left\langle \phi_m^{(ps)}\middle|\phi_n^{(ps)}\right\rangle. \qquad (8)$$

Thus, $\left\langle \Psi_i^{(ref)}\middle|\Psi_j^{(a)}\right\rangle$ can be approximated as

$$\left\langle \Psi_i^{(ref)}\middle|\Psi_j^{(a)}\right\rangle \approx \left\langle \Psi_i^{(ps,ref)}\middle|\Psi_j^{(ps)}\right\rangle + \sum_{m,n}\left\langle \Psi_i^{(ps,ref)}\middle|p_m^{(ps,ref)}\right\rangle q_{mn}\left\langle p_n^{(ps)}\middle|\Psi_j^{(ps)}\right\rangle. \qquad (9)$$

Following the above procedure, we obtained diabatic Hamiltonian matrix and diabatic states.

Taking the MoS$_2$/WS$_2$ TMD system as an example, the diabatization procedure considered two models with different supercell sizes, $6\times 6$ and $9\times 9$ to test whether the results are dependent on the unit size in the electronic structure calculation. Because two models basically give the similar results, we mainly focus on the discussion of the first one. To examine the influence of the numbers of adiabatic VB and CB states included in the diabatization on the final results, we increased the numbers of adiabatic VB and CB states from 10 to 40. As shown in Figure S4, the energies of frontier diabatic states localized on $MoS_2$ and $WS_2$ are consistent for most cases, while the deviation may appear when a very small number of the CB states (for instance only 10 CB states) are involved.

The schemes of the state energy levels for the adiabatic states of the $MoS_2/WS_2$ complex, the adiabatic states of the single $MoS_2$ and $WS_2$ monolayer, and the diabatic states localized either at the $MoS_2$ layer or the $WS_2$ layer of the $MoS_2/WS_2$ complex are shown in Figure 1. For each individual system (the single $MoS_2$ and $WS_2$ monolayer, and the $MoS_2/WS_2$ complex), their zero energy points are taken as the energy of their Femi level. Here 20 VB states and 20 CB states were included in the diabatization. As shown in Figure 1, the $MoS_2/WS_2\_VB@\Gamma$ and the $MoS_2/WS_2\_(VB-1)@\Gamma$ states are the linear combination of the $MoS_2\_VB@\Gamma$ and $WS_2\_VB@\Gamma$ states localized on the $MoS_2$ and $WS_2$ layers respectively. From the electronic density (shown in Figure 2), we can draw the same conclusion. As a contrast, both the $MoS_2/WS_2\_VB@K$ and $MoS_2/WS_2\_(VB-1)@K$ states are localized states, which correspond to the single layer states $WS_2\_VB@K$ and $MoS_2\_VB@K$, respectively. This observation is consistent with the electronic density analysis, as shown in Figures S1 ~ S3 in supporting information (SI). Thus, for both of the $WS_2$ and $MoS_2$ layers, their VB@K states do not interact with any other state of the other layer, when the $MoS_2/WS_2$ heterostructure is formed.

It is also possible to obtain the other diabatic states under the framework of the current projection approach. The linear combination of the $VB@\Sigma$ states of the $MoS_2$ and $WS_2$ layers finally give the $VB@\Sigma$ and $(VB-1)@\Sigma$ states of the $MoS_2/WS_2$ heterostructure. The electronic densities of all relevant states are given in Figures S1 ~ S3 in SI.

The adiabatic-to-diabatic transformation not only assigns the adiabatic states of the whole complex as the linear combination of the localized diabatic states, but also provides the diabatic couplings. Therefore, the current approach is more powerful than the standard analysis of the state components based on the partial density of states (PDOS). As shown in Figure 3, the coupling between the diabatic VB@Γ state located on the $MoS_2$ and $WS_2$ layers is as large as 0.330 eV, while the couplings between VB@K states is only 0.043 eV. The distinct difference between states at Γ and K is in agreement with previous investigations. [5-6, 73-76] At the K point, the interlayer hybridization is weak because of the dominant metal $d_{xy}$ and $d_{x^2-y^2}$ and S $p_x$ and $p_y$ character. The VB is localized on one or the other layer. By contrast, at the Γ point the coupling is much stronger than that at K point and the orbitals are delocalized over both layers because of the mixing of W/Mo $d_{z^2}$ and S $p_z$ orbitals that extend in the $z$ direction (normal to the layer planes). The conclusion from the diabatic couplings is in consistent with the above discussions on the electronic density. Because of the relatively strong coupling, as can be seen from the band structure shown in Figure 4, the energy splitting at Γ point is much larger than that at the K point. It is known that for single layer $MoS_2$ or $WS_2$, the valance band maximum (VBM) locates at K point. [77-81] When the heterostructure is formed, due to the strong interlayer coupling at Γ point, there is an energy splitting as large as 0.67 eV. In this case, the VBM of $MoS_2/WS_2$ residing at Γ point is around 0.2 eV higher than VB states at K point.

Overall, the diabatic VB@Γ state are the most frontier VB states located on $MoS_2$ and $WS_2$, respectively, and their coupling is very strong. Thus, the hole transfer from $MoS_2$ to $WS_2$ (or the reversed electron transfer from $WS_2$ to $MoS_2$) should be ultrafast. When we only take the energy of VB@Γ state of each monolayer (see Figure 1 for their values) and their couplings (~ 0.330 eV) to define a two-state system, the pure electronic dynamics can be obtained. The Rabi-type population oscillation should be expected and the period is around 24.3 fs. This indicates that the ultrafast hole transfer should happen without the inclusion of the electron-phonon couplings. The introduction of the electron-phonon coupling in principle should cause the damping effects, leading to the vanishing of the population oscillation and the suppression of the backward hole transfer. However, considering the extremely fast electronic motion in the pure electronic dynamics, we expect ultrafast hole transfer should still be possible even when vibrational motions are considered. Thus, the pure electronic dynamics based on the current diabatic Hamiltonian in principle capture the zero-order population transfer without the inclusion of the electron-phonon couplings.

On the other hand, the diabatic coupling between two diabatic VB@K states is very weak, resulting in the inefficient direct hole transfer between them. The indirect hole transfer pathway may occur by the involvement of other electronic states. In fact, such pathway were discussed in previous work, which showed that the nonadiabatic transition may take place from the VB@K state to the VB@Γ state through the electron-phonon scattering [82-86]. This opens the possibility of the ultrafast hole transfer. It means that the VB@Γ states are involved during the hole transfer processes, because of the weak couplings between two diabatic VB@K states and strong coupling between two diabatic VB@Γ states

In order to check the performance of the projection method for diabatization, a 9×9 supercell model of $MoS_2$/$WS_2$ heterostructure was also investigated and the results are similar to those of the 6×6 supercell model. As for the details, please see SI.

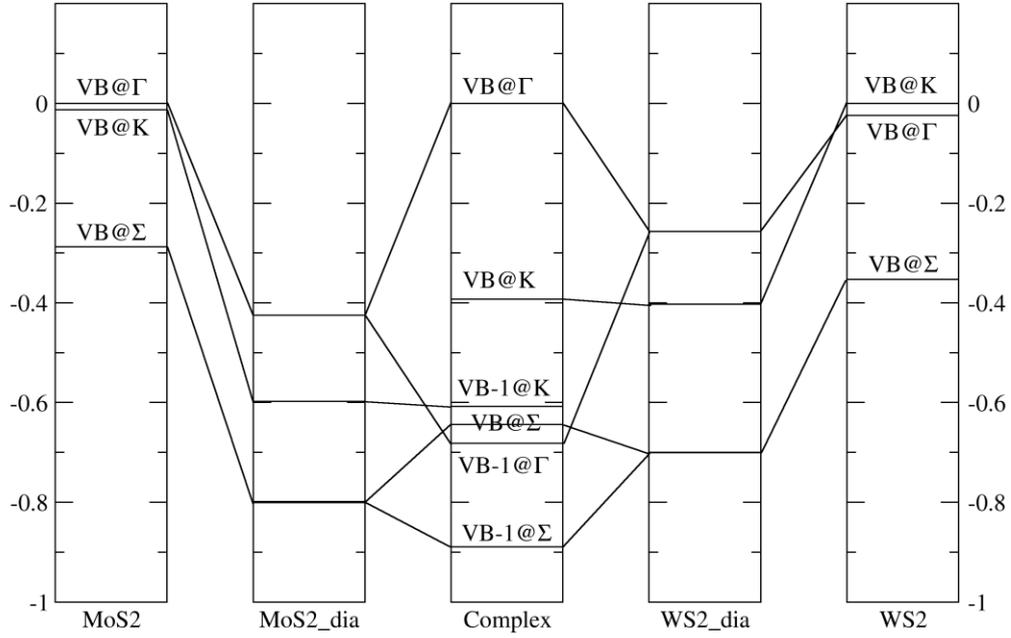

Figure 1. The band-energy-level schemes for adiabatic states in $MoS_2$-monolayer ($MoS_2$), diabatic states located on $MoS_2$ in $MoS_2/WS_2$ heterostructure ($MoS_2\_dia$), adiabatic states in $MoS_2/WS_2$ heterostructure (complex), diabatic states located on $WS_2$ in $MoS_2/WS_2$ heterostructure ($WS_2\_dia$), and adiabatic states in $WS_2$-monolayer ($WS_2$). 6×6 supercell was employed in these systems. For each individual system (the single $MoS_2$ and $WS_2$ monolayer, and the $MoS_2/WS_2$ complex), their zero energy points are taken as the energy of their Femi level.

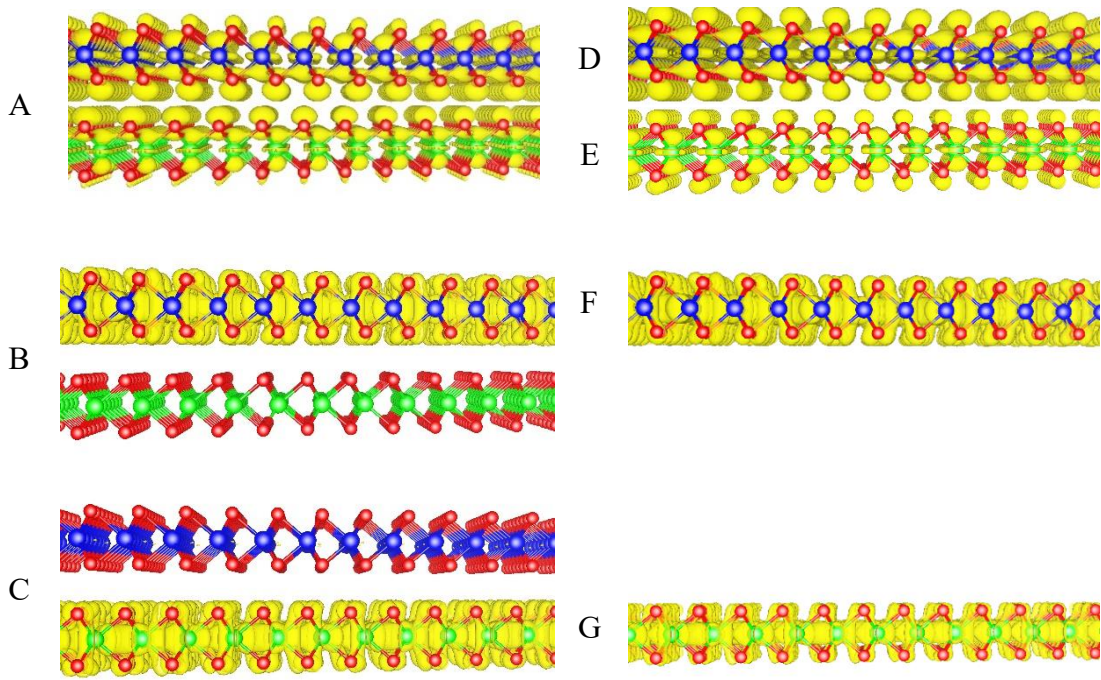

Figure 2. The electron density of diabatic states of $MoS_2/WS_2\_VB@\Gamma$ (A), $MoS_2/WS_2\_VB@K$ (B), $MoS_2/WS_2\_(VB-1)@K$ (C), $WS_2\_VB@\Gamma$ (D), $MoS_2\_VB@\Gamma$ (E), $WS_2\_VB@K$ (F), and $MoS_2\_VB@K$ (G). All the data were obtained with $6\times6$ supercell. The S, Mo and W atoms are represented by red, green and blue balls, respectively.

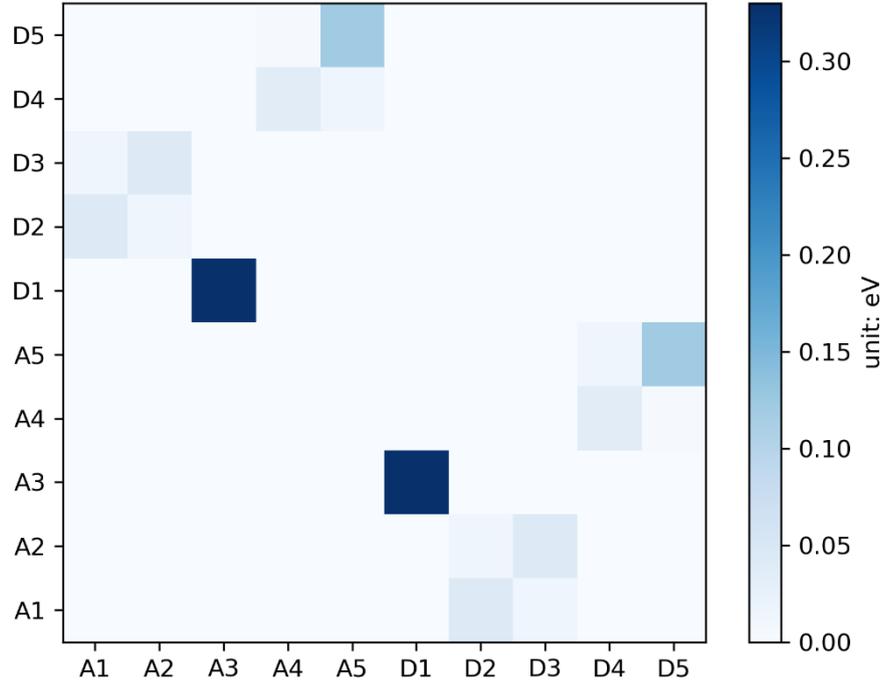

Figure 3. The absolute values of diabatic couplings of $MoS_2/WS_2$ heterostructure with $6×6$ supercell. The diagonal elements in the diabatic matrix are set zero here. The donor states are labelled as D1-D5 according to their decent energy order, while the acceptor states are labelled as A1-A5 in the same way. All labels are: D1 (the diabatic $MoS_2\_VB@\Gamma$ state); D2 and D3 (the diabatic $MoS_2\_VB@K$ states); D4 and D5 (the diabatic $MoS_2\_VB@\Sigma$ states); A1 and A2 (the diabatic $WS_2\_VB@K$ states); A3 (the diabatic $WS_2\_VB@\Gamma$ state); A4 and A5 (the diabatic $WS_2\_VB@\Sigma$ states).

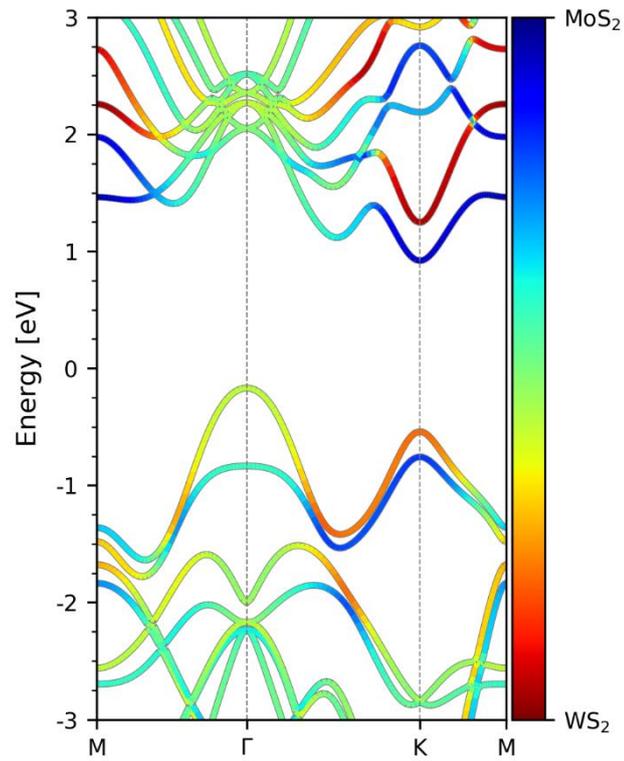

Figure 4. Band structures and orbital spatial distributions of $MoS_2/WS_2$ heterostructure. The momentum-dependent hole localization for different bands is indicated by their coloring according to the color strip.

In the current work, we discuss a practical approach based on the wavefunction projection. Here we wish to give some comments on this method.

We mainly wish to construct the diabatization model for the vdW heterostructure complexes. For such large system, it is very challenging to compute the real excited states. Thus, we still employ a very approximated picture based on the single-particle picture. For instance, in the current $MoS_2/WS_2$ system, the photoinduced hole transfer is simplified as the hole transfer from $MoS_2\_VB$ to $WS_2\_VB$. In this sense, the excitonic effect is missing in our model and the inclusion of it will be an important future effort.

projection-operator approach and block diagonalization of Fock matrix, constrained density functional theory (CDFT), molecular orbital based fragmentation approaches As pointed out in the introduction part, it is certainly possible to construct the diabatic states by CDFT[32, 47-48], the projection operation approach based on the block diagonalization of the Fock matrix [33, 46] and fragment orbital approaches[49-50]. Among them, the diabatization based on the block diagonalization of the Fock matrix is also a kind of "projection method". In such approach, the reference diabatic states are generated by the block diagonalization of sub-block (for instance the donor part) of Fock matrix. The direct combination of this approach and the plane wave basis set is quite challenging. In the current work, we define the reference wavefunction by each layer directly. This provides an easy to perform the wavefunction projection in the framework of the plane wave basis set. Thus; it is easy to employ the current approach in a large amount of theoretical investigations of the similar types of systems based on the calculations with the plane wave basis set. However, also due to the same reason, the current approach is suitable to treat the vdW type systems. When the chemical bonds exist between two parts, the current approach faces problem to define the reference wavefunction in the plane wave basis set.

It is clear that the diabatization results may be dependent on many computational issues, such as diabatization procedure, local basis set v.s. plane wave basis set, the selection of density functionals and so on. Thus; in the future it is certainly interesting to compare the data obtained from the current approaches and other approaches based on the calculations of more benchmark systems. [87] This provide us more deep understanding on the dependence of the diabatization results on the model selection and implementation details.

In summary, we developed a diabatization method for the vdW heterostructure complexes based on the wavefunction projection on the basis of the electronic structure calculations with a plane wave basis set. In this method, we chose the electronic states of individual monolayer structures as the reference diabatic states. By projecting the adiabatic states of the whole complex at any given structure to the reference states, the diabatic Hamiltonian and the diabatic state can be obtained. We chose the vdW $MoS_2/WS_2$ heterostructure complex as a typical example to implement the diabatization approach. The results show that it is possible to obtain the all diabatic states for all frontier VB states. We obtain the transformation between the adiabatic states (frontier VB states of the $MoS_2/WS_2$ heterostructure complex) and the references states (VB states for individual $MoS_2$ and $WS_2$ monolayer). After the construction of adiabatic-to-diabatic transformation, the diabatic Hamiltonian is constructed. This gives the possibility to analyze the components of the adiabatic states of the $MoS_2/WS_2$ heterostructure complex. In addition, the current diabatic approach also provides the electronic coupling (diabatic coupling) between different diabatic localized states. And we obtained the strong coupling at Γ point and weak coupling at K point.

Overall, this wavefunction projection method is a practical approach to construct the diabatic Hamiltonian for the vdW heterostructure complexes. In principle, the current approach may also be employed to treat more general types of extended systems, such as in the cases that the donor and acceptor are connected by chemical bonds. However, in such case, the special attention should be paid to generate the localized orbitals in order to get the correct wavefunctions before and after the formation of the chemical bonds. In addition, it is also important to consider the excitonic states in the extended systems and try to build the diabatic models based on the excitonic states. Such types of works represent a great challenge for the future study.

The current protocol gives not only the energies of the diabatic states, but also their couplings. Thus, the current approach certainly goes beyond the standard PDOS analysis, because it provides the diabatic Hamiltonian for the dynamics study. If we can also estimate the electron-phonon couplings from theoretical approaches, the diabatic Hamiltonian including electronic states and many phonon modes can be constructed. Such diabatic Hamiltonian can be employed to study the photoinduced hole/electron transfer in inorganic photovoltaic material systems. This approach allows us to study photoinduced charge dynamics of 2D stacked complex systems using various advanced nonadiabatic dynamics approaches, including the rigorous full quantum dynamics based on ML-MCTDH [67], the hierarchy equation of motion HEOM [68-70] or different versions of semiclassical approaches[88-90].

## COMPUTATIONAL DETAILS

All electronic-structure calculations have been performed within the Vienna Ab-initio Simulation Package (VASP) [91-93]. Density functional theory (DFT) calculations with the Perdew-Burke-Ernzerh (PBE) functional were used in all calculations. All structures were fully relaxed with a force tolerance of 0.01 eV/Å. The core electrons were treated by the projector augmented wave (PAW) scheme with the pseudo-wavefunctions were expanded in a plane-wave basis set with kinetic energy cutoff 500 eV. The vdW interactions are included in the simulations using the D2 approach. [94] A vacuum region larger than 20 Å is added to avoid the interlayer interactions. All calculations were performed at the Γ point. The pseudo and all-electron wavefunctions were manipulated by using the PIETAS code.[95-96]

## ASSOCIATED CONTENT

**Supporting Information**

The Supporting Information is available free of charge on the website at DOI: xxxxxx. Some additional details of the results, including the frontier states of $MoS_2/WS_2$ heterostructure, $MoS_2$ monolayer and $WS_2$ monolayer with the calculations based on 6×6 supercell or 9×9 supercell, the absolute values of diabatic couplings of $MoS_2/WS_2$ heterostructure, the dependence of results on the inclusion of different numbers of VB states and CBM states in the diabatization.

## AUTHOR INFORMATION

**Corresponding Author**

*Email: zhenggang.lan@gmail.com, zhenggang.lan@m.scnu.edu.cn (Z.L.); renh@upc.edu.cn (H.R.)

**ORCID**

Yu Xie: 0000-0001-8925-6958

Huijuan Sun: 0000-0003-0442-4161

Qijing Zheng: 0000-0003-0022-3442

Jin Zhao: 0000-0003-1346-5280

Hao Ren: 0000-0001-9206-7760

**Notes**

The authors declare no competing financial interest

## ACKNOWLEDGMENTS

This work is supported by NSFC projects (Nos. 21873112, 21673266, 21503248, 21773309, 11620101003 and 11704363). The authors thank the Supercomputing Center, Computer Network Information Center, CAS; National Supercomputing Center in

Shenzhen and National Supercomputing Center in Guangzhou for providing computational resources.# REFERENCES

1. Geim, A. K.; Grigorieva, I. V., Van der Waals heterostructures. *Nature* **2013,** *499* (7459), 419-425.
2. Li, C. L.; Cao, Q.; Wang, F. Z.; Xiao, Y. Q.; Li, Y. B.; Delaunay, J. J.; Zhu, H. W., Engineering graphene and TMDs based van der Waals heterostructures for photovoltaic and photoelectrochemical solar energy conversion. *Chem. Soc. Rev.* **2018,** *47* (13), 4981-5037.
3. Jin, C. H.; Ma, E. Y.; Karni, O.; Regan, E. C.; Wang, F.; Heinz, T. F., Ultrafast dynamics in van der Waals heterostructures. *Nat. Nanotechnol.* **2018,** *13* (11), 994-1003.
4. Wang, G.; Chernikov, A.; Glazov, M. M.; Heinz, T. F.; Marie, X.; Amand, T.; Urbaszek, B., Colloquium: Excitons in atomically thin transition metal dichalcogenides. *Rev. Mod. Phys.* **2018,** *90* (2), 021001.
5. Wang, H.; Bang, J.; Sun, Y.; Liang, L.; West, D.; Meunier, V.; Zhang, S., The role of collective motion in the ultrafast charge transfer in van der Waals heterostructures. *Nat. Commun.* **2016,** *7*, 11504.
6. Zheng, Q.; Saidi, W. A.; Xie, Y.; Lan, Z.; Prezhdo, O. V.; Petek, H.; Zhao, J., Phonon-Assisted Ultrafast Charge Transfer at van der Waals Heterostructure Interface. *Nano Lett.* **2017,** *17* (10), 6435-6442.
7. Zhang, J.; Hong, H.; Lian, C.; Ma, W.; Xu, X.; Zhou, X.; Fu, H.; Liu, K.; Meng, S., Interlayer-State-Coupling Dependent Ultrafast Charge Transfer in MoS2/WS2 Bilayers. *Advanced Science* **2017,** *4* (9), 1700086.
8. Zhang, J.; Hong, H.; Zhang, J.; Fu, H.; You, P.; Lischner, J.; Liu, K.; Kaxiras, E.; Meng, S., New Pathway for Hot Electron Relaxation in Two-Dimensional Heterostructures. *Nano Lett.* **2018,** *18* (9), 6057-6063.
9. Ji, Z.; Hong, H.; Zhang, J.; Zhang, Q.; Huang, W.; Cao, T.; Qiao, R.; Liu, C.; Liang, J.; Jin, C.; Jiao, L.; Shi, K.; Meng, S.; Liu, K., Robust Stacking-Independent Ultrafast Charge Transfer in MoS2/WS2 Bilayers. *ACS Nano* **2017,** *11* (12), 12020-12026.
10. Long, R.; Prezhdo, O. V., Quantum Coherence Facilitates Efficient Charge Separation at a MoS2/MoSe2 van der Waals Junction. *Nano Lett.* **2016,** *16* (3), 1996-2003.
11. Liang, Y.; Li, J.; Jin, H.; Huang, B.; Dai, Y., Photoexcitation Dynamics in Janus-MoSSe/WSe2 Heterobilayers: Ab Initio Time-Domain Study. *J. Phys. Chem. Lett.* **2018,** *9* (11), 2797-2802.
12. Hong, X. P.; Kim, J.; Shi, S. F.; Zhang, Y.; Jin, C. H.; Sun, Y. H.; Tongay, S.; Wu, J. Q.; Zhang, Y. F.; Wang, F., Ultrafast charge transfer in atomically thin MoS2/WS2 heterostructures. *Nat. Nanotechnol.* **2014,** *9* (9), 682-686.22